# Nanobubbles, pristine emulsions, high ionic strength electrokinetics – paradoxes of Colloid and Interface Science.

*Andrei Dukhin[1], Renliang Xu[2], Darrell Velegol[3]*

[1] adukhin@dispersion.com, Dispersion Technology Inc, 364 Adams Street, Bedford Hills, NY 10507

[2] renliang.xu@yahoo.com, 13084 NW 13 Street, Pembroke Pines, FL 33028

[3] velegol@psu.edu, Department of Chemical Engineering, Penn State University, *CBEB 225, University Park PA 16802*

- *Correspondence*: Dr. Andrei Dukhin

## Abstract

There are three paradoxes in the modern Colloid and Interface Science supported by large bulk of experimental evidence and contradicting classical theoretical models:

- Electrolikinetics at high ionic strength;
- Nanobubbles having life span on scale of days and weeks without any surface stabilization;
- Pristine emulsions having life span on scale of days and weeks without any surface stabilization.

We overview many dozens of experimental papers by broad spectrum of scientific groups from many countries. We consider this vast experimental data as unambiguous evidence that these phenomena exist. On other hand, classical theoretical models deny such possibility.

This contradiction between experiment and theory justifies introduction of new theoretical models. We overview these models. It turns out that there is one common feature between most promising of them – assumption regarding structured water layer at hydrophobic interface. That is why we combine these three phenomena in this review.

There is extensive literature on century old idea of the structured water layer, experimental and theoretical. We overview this literature, which provides convincing support to this hypothesis.

We discuss existing theoretical models that incorporate the structured interfacial water layer for the successful explanation of these phenomena in more detail. Theoretical model of electrokinetics at high ionic strength was developed several decades ago. Theoretical model explaining paradoxical longevity of nanobubbles and pristine emulsions by interaction between structured water layer and electric double layer is more recent.

## Introduction

There are three phenomena in the modern Colloid and Interface Science that contradict classical theories and should not exist at all. This list includes electrokinetics at high ionic strength, nanobubbles and pristine emulsions prepared with no surfactants and having longevity on scale of days and weeks. All these paradoxical phenomena have common roots according to the review presented here. That is why we combined them together. We start with electrokinetics at high ionic strength because it is much simpler than other two effects and there is already well established theory explain it.

Electrokinetics at high ionic strength exceeding 1 mol/dm$^3$ should not exist according to Gouy-Chapmann (GC) theory [1,2] because Debye length [3] becomes comparable with ion size. Nevertheless, this phenomenon was observed by 5 independent groups using 5 different measuring methods [4-11].

There were attempts to explain these experiments with ion sizes, image forces, and entropy factors that are ignored in the classical EDL theory, similarly as this was done with the Bickerman-Friese EDL model [12-16]. However, Lyklema [17] concluded, based on numerical calculations conducted by Carnie and Torrie [18] that all discrete theories lead to the reduction of the surface potential for a given surface charge compared to the classic GC theory. Consequently, they cannot explain excessive surface potential at high ionic strength.

The alternative theory was introduced by Mancui and Ruckenstein [19] who employed hypothesis of the ***structured water interfacial layer***. They succeeded in explaining experimental data. This approach was extended recently by Arias in the paper [20]. Similar approach was used by Yaroschuk to explain observed shift of iso-electric point at high ionic strength [21].

We overview and discuss these experimental and theoretical studies in detail in the section that is dedicated to this phenomenon.

The hypothesis of the ***structured water interfacial layer*** was suggested also for explaining existence of surfactant-free "pristine" emulsions by Eastoe and Ellis [22] and Pashley [23]. The term "**pristine**" was introduced by Beattie and Djerdjev [24] for stressing difference of these emulsions [25-49] with Pickering emulsions [50]. Pristine emulsions should not exist according to traditional paradigm in Colloid and Interface Science, which states that "… *droplets or bubbles are not stable without an adsorbed layer of a surface-active component (surfactant, polymer, polyelectrolyte, protein)…*[51]. Nevertheless, there are at least 27 papers published about these emulsions, which we overview in the section dedicated to this phenomenon.

The droplets of these emulsions are charged. This charge occurs due to spontaneous adsorption of hydroxyl ions on water with non-polar oil interface, as was shown very convincingly by Marinova et [25] and many other authors. There is also possibility of contribution by ions associated with $CO_2$ adsorption, as discussed in the papers [43-46]. Such open-to-air pristine emulsions could be considered as aqueous carbonated systems [52] with carbonic acid ions adsorbed at the interface.

This charge stabilizes emulsions against coalescence according to DLVO theory [53-55]. However, these emulsions exhibit some features that cannot be explained within framework of DLVO theory. Here is citation from the paper by Pashley [23]: "…*The failure of the DLVO theory to predict behavior oil-in-water emulsions raises the issue of whether there might be additional forces involved…*". Then, the authors proceed by introducing so-called "hydrophobic interactions" [56] that are associated with ***structured water interfacial layer***. The concept of the "hydrophobic interactions" requires rather thick structured layer that penetrates the bulk up to 10 nm. There is another principle that could account for additional stability factor and emulsification at much shorter, much more reasonable structured layer, with thickness below 0.5nm. It was suggested in papers [48,49,57]. We discuss this in detail in the section dedicated to pristine emulsions.

This new recent theory has been developed initially for explaining longevity of nanobubbles – another class of the systems that should not exist according to the IUPAC report [51]. Nanobubbles are gas-filled cavities in liquid, usually in water. Their diameters are in submicron range, mostly 100-500 nanometers The well-known theory by Epstein and Plesset [58,59] estimates lifespan of bare nanobubbles in water around 0.001 sec. It is so short because Young-Laplace excessive pressure [60,61] causes diffusional outfluxes of gas and the bubbles quickly dissolve into the water bulk. Despite this, existence of surfactant-free nanobubbles with life span on scale of days and even weeks has been established by many independent groups [62-85]. Recent review on nanobubbles by Docar et al [82] has 159 references.

There have been several hypotheses suggested for resolving this contradiction between classical theory and experiment, overviewed in paper [82].

The leading one is associated with electrostatic effect of the electric surface charge [83-85]. It is well known that bubbles are getting charged due to adsorption of hydroxyl ions, similarly to emulsions [86-92]. In essence, proponents of this hypothesis attempted to extend and modify well-known Onsager-Samaras theory [93] describing effect of EDL on the surface tension for nanobubbles. Authors of the paper [82] conducted systematic review of such approach and concluded that effect of EDL on the surface tension is too weak for explain nanobubbles stability. This agrees with Onsager-Samaras theory, which predicts only a few percents corrections to the surface tension from EDL.

At the same time, there is ample experimental evidence indicating that electric surface charge and associated EDL play important role in nanobubbles stability and structure. There is another mechanism that allows for much stronger impact of EDL on the nanobubbles. It was suggested in the papers [48,49,57]. In principle it offers full balance between normal and lateral stresses at the nanobubble interface. It is discussed in detail in the section below. This mechanism requires existence of the ***structured water interfacial layer*** at the nanobubble interface.

Existence if such layer is related to the other hypothesis that is invoked for explaining nanobubbles stability - hydrogen bonds [62,66,94,95]. Hydrophobic substrate, either oil or gas, disrupts water hydrogen bonds in its vicinity. Such disruption leads to reorientation of water molecules. The existence of structured water on hydrophobic interfaces has been confirmed not only experimentally but by computer modeling. Here is quotation from recent one by Walker-Gibbons et al [96]: "…*Careful examination and comparison of hydrogen bonding in layers of interfacial water approaching the surface strongly suggest that the excess interfacial potential indeed arises from the broken hydrogen bonding symmetry in the presence of an interface. Importantly, we find that the net orientational behavior of water is found to be similar across a range of surface chemistries…"*.

This layer of oriented water molecules is thin, 0.2-0.4 nm in thickness. It encompasses 1-2 monolayers of water molecules. All these molecules possess significant dipole moments. These parallel oriented dipole moments repel each other and in principle must destabilize interface even more, in contrary to some papers considering this structured water layer as mechanical obstacle for diffusions outflows fluxes. However, theoretical model suggested in papers [48,49,57] managed to resolve this contradiction.

There is one simple way to verify validity of this idea, and it has been accomplished in the paper [81]. The low and non-polar liquids have much smaller molecular dipole moments. Terefore, nanobubbles in such liquids should be much less stable if these surface dipoles are important. Authors of the paper [81] observed indeed increase in bubble size and dramatic reduction of longevity compared to water. They concluded that liquid polarity and related molecular dipole moments are important for nanobubbles stability. However, the exact mechanism was still uncertain. It was suggested eventually in the papers [48,49,57] and we review in here in detail.

S***tructured water interfacial layer*** is a common element that is present and apparently plays important role in all three discussed paradoxical phenomena. That is why we start detail sections from overview of papers dedicated to its studies.

There is one more factor that could be important for all phenomena – impurities. It deserves a special comment. In principle, all these effects and experiments could be affected by traces of organic substances that could modify interfaces and create artifacts. That is why there were often suggestions to designate impurities as origins of all these paradoxes.

All authors of the papers cited here realized this danger. There have been different ways developed for minimizing impurities impact.

First, sophisticated cleaning procedures were employed for reducing potential contribution of impurities in oil and water phases to small fractions of percent.

Second, some creative ways of samples preparations were invented.

For instance, Marinova et al [25] managed to eliminate homogenizer completely by heating up to $60^0$ C oil and water mixture. Oil dissolves slightly in water at this higher temperature. Then, when this mixture cools down, dissolved oil forms spontaneously droplets with sizes 100-200 nm.

Another way to achieve such spontaneous emulsification is repeated freeze-pump-thaw treatments described by Pashley [23,32].

Increasing volume fraction of oil phase in oil-in-water emulsion leads also to reducing relative role of impurities potentially contained in the dispersion medium due to increasing total interfacial area.

There is one more argument that eliminates impurities as important factor – high reproducibility of the measured data. And this is widely recognized in the field. For example, here is the quote from one the published papers [77] :"*…if contaminations, owing to inflow of pollutants in the air, desorption of organic or inorganic substances had occurred during the experiment, the results of zeta potential measurement would not show a uniform value, and the standard deviation would have been large…*". Meantime, zeta potential and nanobubble size can be measured quite reproducibly, and they exhibit clear correlation according to the papers [78-80].

Similar arguments could be found in many other papers.

## Structured interfacial water layer at hydrophobic interfaces.

The initial idea of Hardy [97] in 1915 about water structuring at interface was further promoted 30 years later in 1945 by Frank and Evans [98]. These authors introduced an idea of water structuring when it gets into contact with non-polar substances. They wrote: "*…When a rare gas atom or a non-polar molecule dissolve in water at room temperature it modifies the water* *structure in the direction of greater crystallinity – the water, so to speak, builds a microscopic iceberg around it…".*

Independently Bickerman studying electric double layer (EDL) at electrodes in 1942 realized role of water molecules dipole moments [12]. In particular he understood that EDL inhomogeneous field exerts force on the water dipoles. Here is what he wrote: *"...An ion moving in an aqueous solution permanently exchanges places with water dipoles. If this movement occurs in an inhomogeneous electric field a displacement of dipoles generally requires a (positive or negative) work...*" [12]. Essentially, he stresses the importance of dipoles of water molecules on events that occur in EDL.

The next step on this path was made by Bockris with co-authors in 1963 [99], again studying EDL of electrodes. They declared that water molecules in at least the first monolayer of EDL must be oriented due to the high electric field that affects their dipoles. This gives rise to the idea of *structured water layer* at the interface inside of EDL.

This opinion was shared and extended to colloidal systems by Hunter in his famous book on Zeta potential in 1986 [100]. Here is the quote from this book: "…*the high electric field near the wall (~$10^6$-$10^8$ V/m) must produce some ordering of solvent dipoles, especially for water*…". He stated that this model is supported by many experimental and theoretical studies, starting from 1928.

The first direct experimental discovery of the structured interfacial layer was achieved by Pashley and Israelachvili in 1984 using ***Atomic force microscopy*** [101]. Here is Lyklema's comment of this study "…*the maxima must reflect the layer-wise ordering of water near the surface with periodicity of about 0.24 nm, which is close to diameter of water molecule*…"[17].

More recently several advanced spectroscopic techniques were involved for studying water at hydrophobic interfaces.

***Raman excited fluorescence*** (SREF) was employed by Xiong et al in 2020 [102] for study of the aqueous microdroplets interfaces. Authors discovered that strong electric fields at interface align probe dipoles with respect to the interface and seems that water molecules with large dipole moments should do the same.

***X-ray pair distribution function*** analysis was used by Zobel at al in 2015 [103] for studying polar and non-polar solvents restructuring around nanoparticles. They established that *"….layers of enhanced order exist with a thickness influenced by the molecule size and up to 2 nanometers beyond the nanoparticle surface…"*.

***Vibrational sum frequency spectroscopy*** (VSFS) was used by Tyrode and Lijeblad in 2013 [104] for studying the structure of water in contact to hydrophobically modified fused silica surfaces. Here is their conclusion: "*…the results imply that the structure of water in the most ordered monolayers is not much affected beyond the first layer of water molecules, with bulk isotropic properties becoming apparent already at sub-nanometer distances from the surface…*"

There is another approach to studying surface water properties that can be characterized as "***water in confinement***".

One of early examples of such study is measurement of the water ***thermal expansion*** in silica gel with pores around 5 nm in size conducted by Derjaguin et al in 1986 [105]. They established that *"….the detected specific properties of the boundary water can be ascribed to its layers having a thickness of the order of some scores of Angstrom units…"*. They also claimed that the difference attributed to the structured water layer disappears at temperatures above $70^0$ C. This opens opportunity for additional tests of theoretical models discussed below.

The more recent example is measurements of the ***shear and longitudinal sound speed*** of the water in confinement described in the paper [106].

There are many more studies on this subject. There the extensive review by Wang published in 2026 [107]. This review contains ***262*** references to the studies of the water in membranes, nano tubes, zeolites and other systems having opening with sizes on nanometers scale. Most of the studies apply various spectroscopic techniques for studying water trapped in such small cavities. Here are some citations from this review: "*…directly visualized ordered bilayers of water even at graphitic surfaces, revealing that interfacial structuring can coexist with hydrophobicity….*"; "*….water near hydrophobic surfaces carries a net negative interfacial potential, associated with preferential orientation of O–H bonds and entropic stabilization…*"; "*…the O–H dipoles at a hydrophobic alkyl-terminated surface are oriented…*".

It seems that there is enough evidence now to conclude that such *structured water layer* affecting ions solubilities exists indeed. It could affect the entire EDL and related electrokinetic effects.

## Electrokinetics at high ionic strength

We found publications by 5 independent groups using 5 different measuring techniques for studying electrokinetics at high ionic strength in both particulates and porous materials. These results unequivocally prove the existence of EDL at high ionic strength. We combined all main results illustrating measurable zeta potential on a single figure – Figure 1.

The first study was conducted with electroacoustic method in the ESA mode [108-111]. The authors, Johnson, Scales and Healy, reported zeta-potential of alumina in the range of 20-30 mV at 1 mol/dm$^3$ ionic strength of several electrolytes, as described in Figure 1.a [4].

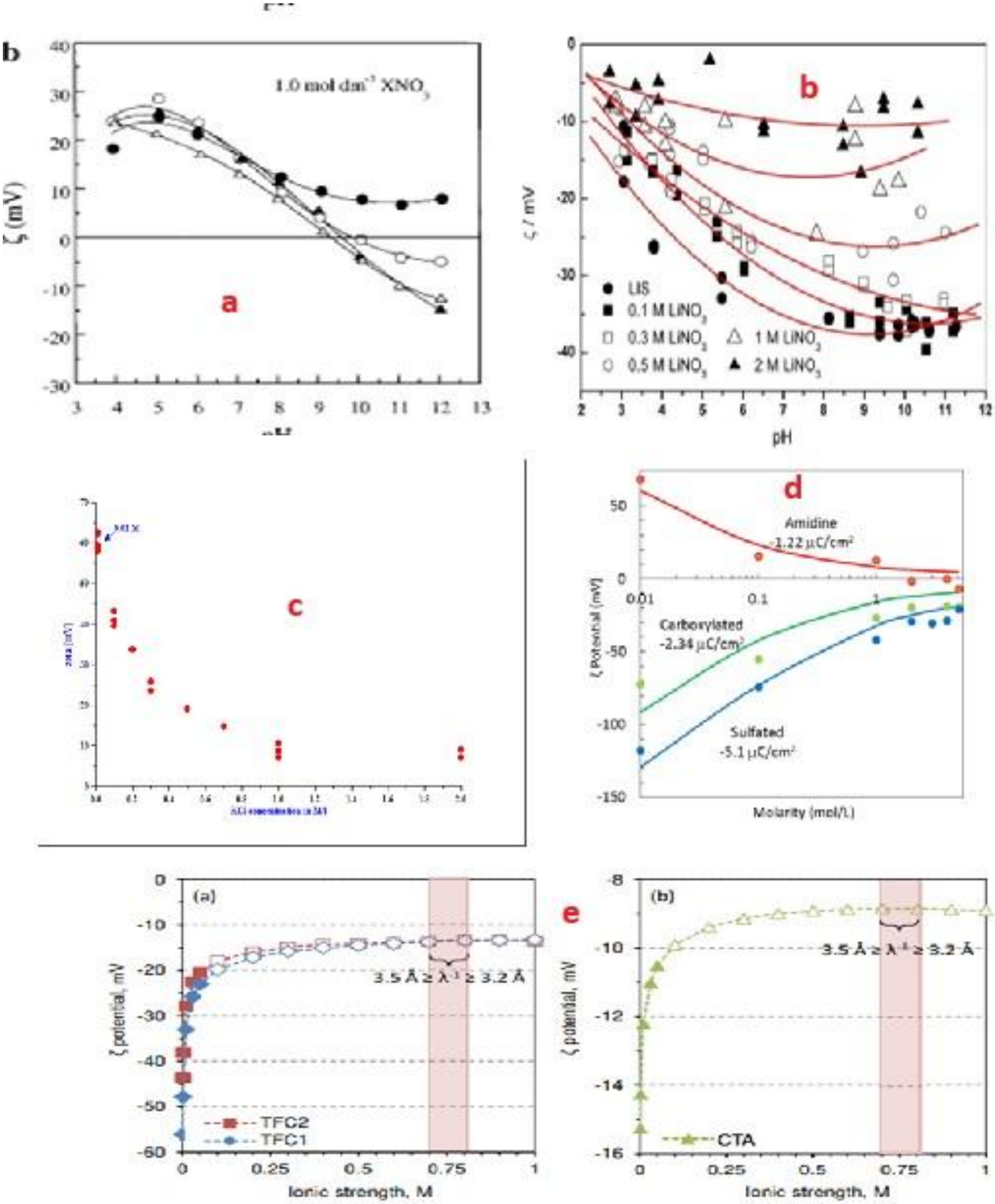
1.0 mol dm⁻³ XNO₃
ζ (mV)
a
b
ζ / mV
pH
LiIS
0.1 M LiNO₃
0.3 M LiNO₃
0.5 M LiNO₃
1 M LiNO₃
2 M LiNO₃
c
d
Amidine
1.22 μC/cm²
Carboxylated
-2.34 μC/cm²
Sulfated
-5.1 μC/cm²
ζ Potential (mV)
Molarity (mol/L)
(a)
3.5 Å ≥ λ⁻¹ ≥ 3.2 Å
ζ potential, mV
TFC2
TFC1
Ionic strength, M
e
(b)
CTA

**Figure 1. (a) The zeta potential behavior of α-alumina as a function of both monovalent cation type and pH. The electrolyte concentration is 1.0 mol/dm$^3$ XNO3. The alumina volume fraction is 0.020 in all cases. Key: (●) X= Li; (○) X= Na; (▲) X =K; (Δ) X= Cs. (Reproduced from [4]); (b) The zeta potentials of montmorillonite in $LiNO_3$ solutions. (Reproduced from [7]); (c) Zeta-potential of alumina AKP30 at 3 %vl versus KCl concentration. (Reproduced from [9]); (d) Zeta potential plotted against NaCl salt concentration on a logarithmic scale, for each of the 3 particles tested (circles) along with chi-squared fits for each particle (solid lines). The number below each particle name is the best fit surface charge value. (Reproduced from [10]; (e) values of zeta potential calculated from measured streaming potential for (a) TFC1 andTFC2, and (b) CTA membranes as a function of electrolyte (KCl) ionic strength. (Reproduced from [11]).**

The next group by Kosmulski, Rosenholm with co-authors published many papers on this subject [4-7]. They studied a wide variety of particles including alumina, anatase, zirconia, kaolin, montmorillonite, titania, silica, indium and niobium oxides, goethite, cerium dioxide. They used a wide variety of electrolytes and 3 different measurement technologies: dynamic light scattering, electroacoustics in the ESA mode, and electroacoustics in the CVI mode [110,111]. We reproduce here just one figure (Figure 3(b)) from these extensive studies.

Here are some conclusions from the authors [4]:

"*…The results obtained by means of three different instruments based on different principles are qualitatively consistent. Thus, the observed phenomena are likely to be real effects rather than instrument artifacts……The present study confirmed that the ζ potentials in 1 M 1-1 electrolyte solutions can assume values in excess of ±20 mV …*"

"*…The alkali metal cations, especially Li and Na, adsorb specifically from concentrated solutions of their halides, nitrates and chlorates, on metal oxides. This results in a shift in the IEP of these materials to high pH values and in abnormally high positive values of the zeta potential…*"

Electroacoustic method in the CVI mode was employed by Dukhin et al [9]. They studied concentrated dispersion of alumina using. They conducted zeta potential measurements versus ionic strength instead of running pH-titration. Figure 1(c) shows their result. It is seen that zeta-potential reaches a stable value around 10 mV at high ionic strength instead of going to 0 as it is predicted by the GC model.

The group of Velegol from Penn State University [10] found ingenious ways of modifying electrophoretic method for making it suitable for high ionic strength. They used microscopic video recording of electrophoretic motion of latex particles in various ionic strengths up to 5 mol/dm$^3$ . The applied electric field was in AC mode at 300 Hz. To minimize heat and convection effects, measurement time was reduced to 2 s. Figure 1(d) represents their main results. They also observed that zeta potential retains finite values at very high ionic strength for negatively charged particles. In the case of positively charged ones it goes to zero.

The authors concluded that these observations indicate that the detailed structure of the EDL at high salt depends on the chemical identity of all ions present in the electrolyte. By contrast, the zeta potential at low salt depends only on the ionic strength and perhaps the identity of some potential determining ions (e.g., hydroxide).

There is also a study in porous material by Coday, Luxbacher et al [11]. They use the streaming potential method at high ionic strength. They measured zeta potential of three different membranes as a function of ionic strength up to 1 M solutions. Results are shown in Figure 1(e). It is seen that an absolute value of zeta potential reaches saturation level around 10 mV similarly to the alumina study in the paper [9].

These five experimental studies prove unambiguously the existence of electrokinetics at high ionic strength. This requires electric charges separation at interface. The presence of the ***structured water layer*** at interface offers following mechanism of such separation, which was discussed in the papers [19,20,21].

From a phenomenological viewpoint, the cations and anions may have quite different ability to penetrate into the interfacial water layer due to differences in solvating. This in turn might lead to differences in their concentration within this layer and consequently, to separation of electric charges. The excess charge in the structured water layer would in turn generate an electric field that then builds a screening diffuse layer in the bulk. Figure 2 illustrates this model.

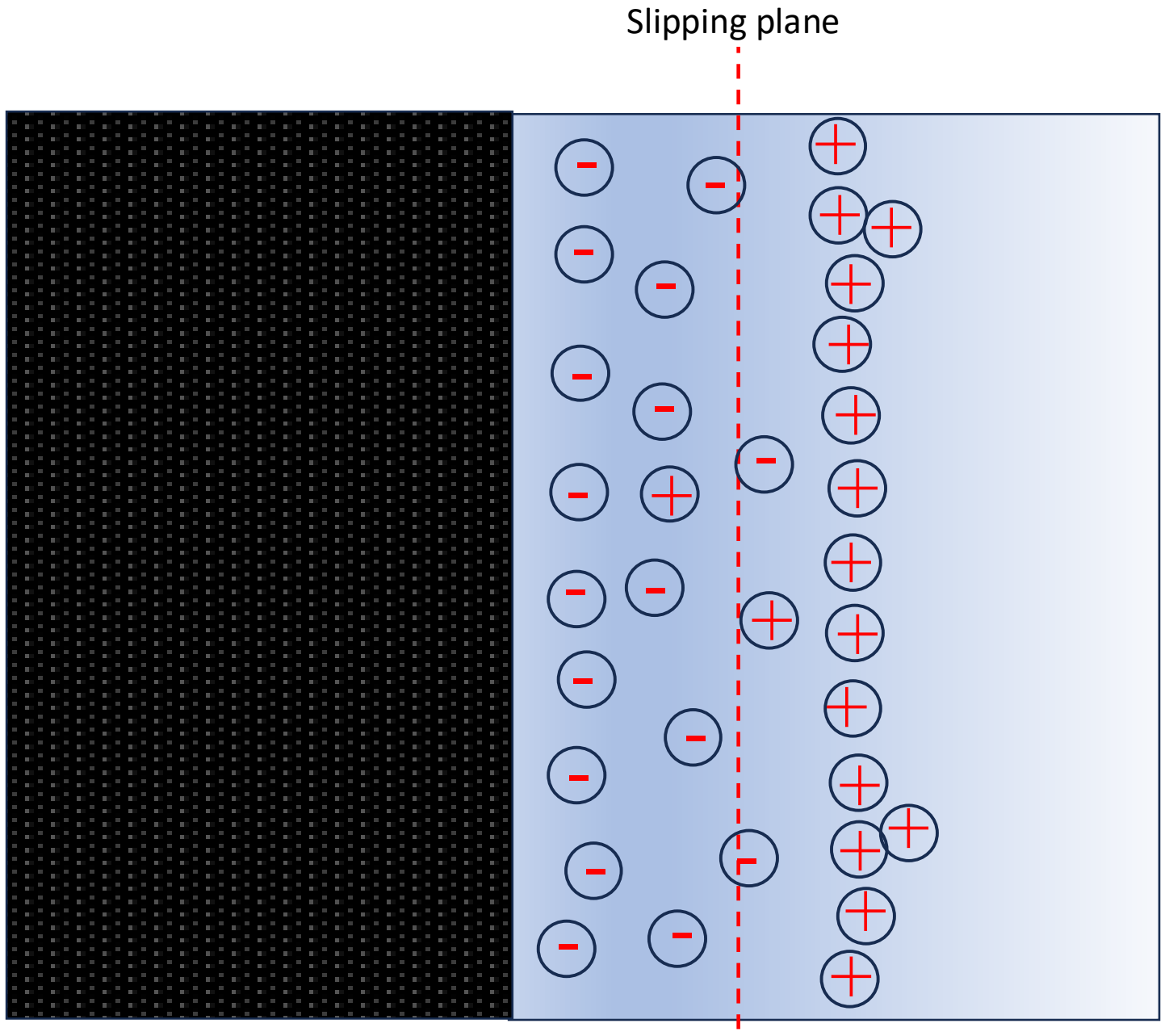

**Figure 2. Illustration of electric charges separation and formation of EDL due to structured water layer.**

Interfacial charge in this model is associated with the water phase. Structural peculiarities of the surface water layer diminish towards the bulk. Deviations in ions concentrations induced by this structure would decline correspondingly. For instance, if water interfacial structure allows higher concentration of anions and lower concentration of cations, as on Figure 2, then there would be excessive negative charge next to the interface. The density of this interfacial charge would decline with the structure. This interfacial charge electro-statically attracts counterions (positive on Figure 2), same as the traditional surface charge.

At high ionic strength, this layer of counterions becomes flat, as shown in Figure 2 because Debye length becomes equal to ion size. It looks like diffuse layer and surface charge layer switch sides compared to the GC model. The reason for the diffuse structure of the internal layer differs as well. It is not thermal motion as in the GC model, but the diffuse structure of surfaces forces limiting ions solubility.

If the outer layer of counter ions lies outside of the slipping plain, it becomes hydrodynamically mobile. An external electric field could move it, giving rise to electroosmotic slip and consequently, electrokinetic phenomena.

The qualitative theory was developed by Mancui and Ruckenstein EDL [19]. They attempted to fit experimental data on zeta potential of bubbles, as the mechanism of charge formation due adsorption of $OH^{-1}$ on water-gas interface is well known. They assumed the thickness of this structured water layer $L_{str}$ to be 0.4 nm and made assumption about different solubilities of cations and anions in that layer. They managed to achieve a reasonably good fit and concluded that one important consequence of the present hydration model is that the predicted surface potential does not vanish at high ionic strength because of the cut-off distance for cations. Since there is no screening in the cation depletion region, the surface potential $\psi^s$ at large electrolyte concentration according to their model is roughly given by:

$$\Psi^s(I \to \infty) = \frac{L_{str}\sigma^s}{\varepsilon_0 \varepsilon_m} \qquad (1)$$

Where, I is ionic strength, $\sigma^s$ is surface charge density, $\varepsilon_0$ and $\varepsilon_m$ are dielectric permittivity of the vacuum and the liquid.

They also concluded that the asymmetric distribution of ions in the structured layer generates a surface potential, which increases with increasing electrolyte concentration.

There is more sophisticated version of such electrokinetic theory created recently by Arias [20].

There is one more aspect of the high ionic strength electrokinetics that can be resolved with existence of ***structured water layer***. It is a shift in iso-electric point towards higher pH rage with increasing ionic strength, as was reported by Kosmulski and Rosenholm group [4]. The nature of this effect was revealed by Yaroshchuk [21]. His model assumes "*…that the standard solvation energies of ions at the vicinity of the interface differs from those in the bulk due to the changes in the solvent structure…*". This is the same as in the previously mentioned theoretical papers. He developed quantitative theory that explains experimentally observed shifts.

There was attempt to verify relationship between high ionic strength electrokinetcs and structured interfacial water directly in the paper [9]. Authors used surfactant with hope to move slipping plane away from the interface and beyond reach of the structured water layer. If succeed this method would eliminate mobility of the outer electric charges in the EDL shown in Figure 4 and electrokinetic effects must cease to exist in such surfactant modified dispersion. Unfortunately, this attempt failed because high ionic strength prevents surfactant adsorption.

There is another approach for verifying this theory – conducting similar experiments at elevated temperature. The water structure at interface should decline with increasing temperature. Correspondingly zeta potential in the solution with high ionic strength must gradually decrease with increasing temperature. We do not know about such experiments yet.

It seems that there is enough evidence now to conclude that such ***structured water layer*** affecting ions solubilities exists indeed. It could affect the entire EDL and related electrokinetic effects.

## Nanobubbles.

Experimental data on the exceptional nanobubbles longevity is so vast that it is becoming impossible to ignore this paradox and forces us to accept nanobubbles stability and longevity as a fact. Then we are facing several theoretical questions that underline the conflict between this experimental fact and classical model of liquid-gas interface.

1. Young-Laplace excessive pressure inside of the bubbles causes diffusional outflow of the gas that is supposed to limit nanobubble lifetime to fraction of millisecond. What factor stops it?
2. Even if we would find such factor, what does then balance the surface tension at the presumably stable nanobubbles interface?
3. Why is the reported size range of stable nanobubbles specifically around 100-500 nm?

There is a theory that provides answers to all these questions [48,49,57]. It has been cited already in papers [20, 35,82,112-116]. This theory considers nanobubbles as a gas cavity covered with two interfacial layers: ***Electric double layer*** and ***Structured water layer***, as it is shown on Figure 2.

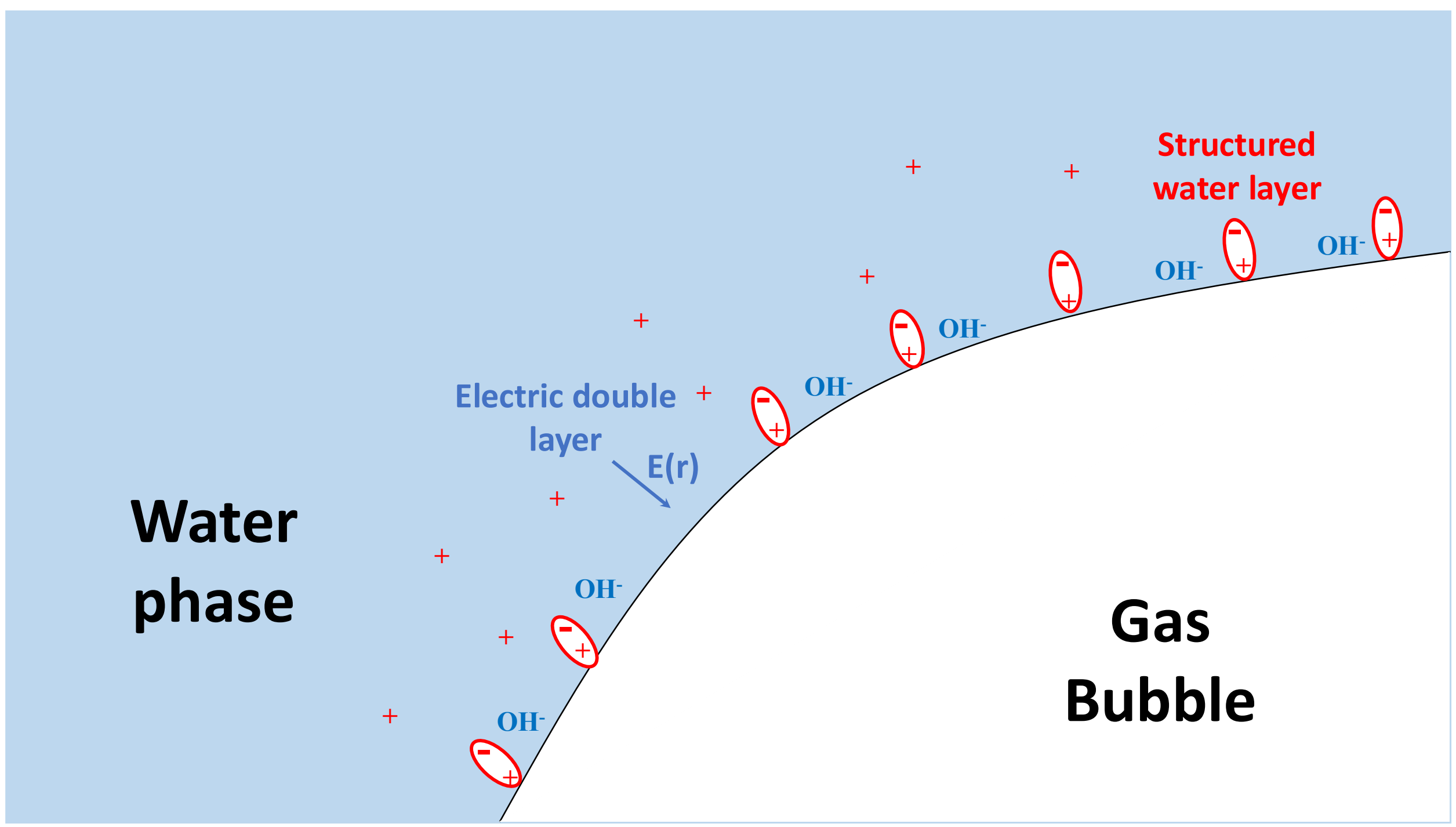

**Figure 2. Illustration of nanobubbles covered with two interfacial layers: Electric double layer and Structured water layer.**

There are serious supporting arguments discussed in the Introduction of such a model of nanobubble.

Both these interfacial layers cannot resolve longevity paradox when taken separately. However, it turns out that they interact with each other and this interaction produces desirable effect. The mechanism of this interaction is shown in Figure 3.

Electric field of EDL is inhomogeneous in the sense that electric field strength E(r) is higher closer to the interface and declines with distance from the interface r. This inhomogeneity is shown by two blue arrows next to the oriented dipole moment of the water molecule at interface. The EDL electric field exerts electrostatic forces on the electric charges of the dipole $F^{\pm}$, which are illustrated as green arrows in Figure 3. The force acting on the positive charge is larger because it is closer to the interface and experience higher electric field strength. In addition, these forces are oppositely directed because charges are of opposite signs. Resulting force acting on the whole water molecule is directed toward interface and is shown as red arrow. It was named as "dielectrostatic force" in the paper [57], following analogy with dielectrophoresis [117]. This force competes with Young-Laplace pressure and in principle could disrupt diffusional gas out fluxes.

There is an expression for the pressure that this force exerts to a bubble $\Delta P_{wd}$ for further comparison with the Young-Laplace pressure derived in the paper [57]:

$$\Delta P_{wd} = \frac{F_{w-bubble}}{S} = d_w \kappa^2 \zeta * 1000 * L_{stl} * 55.5 * N_A \qquad (2)$$

where $d_w = 6.17*10^{-30}$ C*m is the water molecule dipole moment, a is the bubble size, r is distance from the bubble center, $\kappa^{-1}$ is Debye length, and $\kappa^2 \approx (1/9)*I*10^{20}$ [1/m$^2$], $I$ is ionic strength in [mol/dm$^3$], ζ is zeta potential, 55.5 is moles of water molecules per dm$^3$, and $N_A$ is the Avogadro number, $L_{stl}$ is the thickness of the structured water layer.

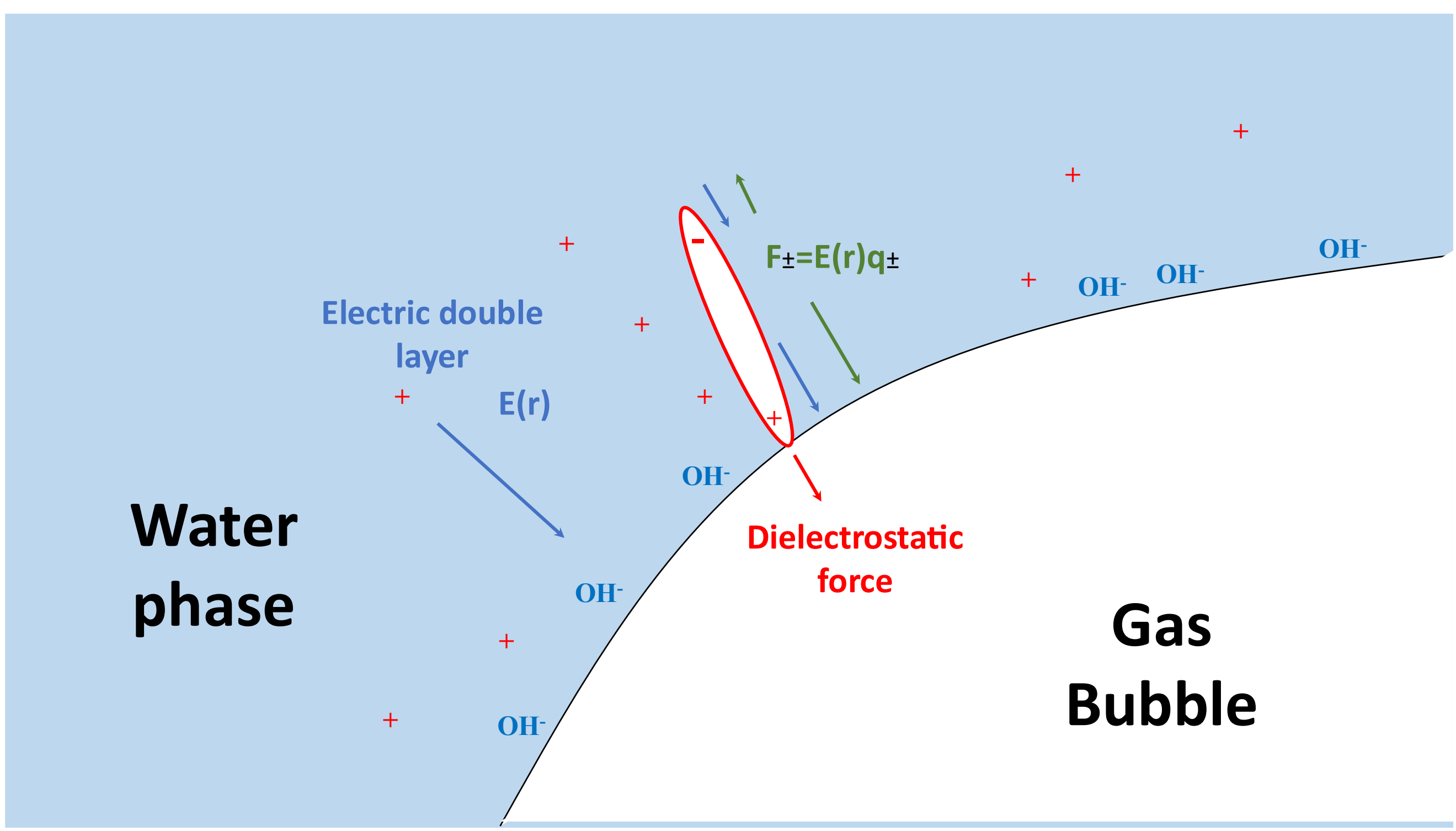

**Figure 3. Illustration of interaction between EDL inhomogeneous electric field E(r) and the dipole moment of the oriented water molecules in the structured layer, leading to dielectrostatic force.**

The value of the dielectrostatic force pressure must be compared with the well-established expression for the Young-Laplace equation $\Delta P_{YL}$ between bubble's interior and exterior:

$$\Delta P_{YL} = \frac{2\gamma}{a} \qquad (3)$$

where $\gamma$ is the surface tension which equals 0.072 N/m for water-gas interface, a is the bubble radius.

Comparison of these pressures for the parameters that correspond to real experiments revealed that they are very close. Basically, their exact balance

$$\Delta P_{YL} = \Delta P_{wd} \quad (4)$$

determines a stable nanobubble size $a_{stable}$ :

$$a_{stable} = \frac{2\gamma}{\zeta * \kappa^2 * d_w * L_{stl} * N_A * 55500} \qquad (5)$$

This equation could be verified using experimental data from the papers that contains results of measurements for both nanobubbles sizes and zeta potentials. We selected data from the paper [78] for such test and reproduce graph from this paper here as Figure 4. Then we assume value of the structured water layer thickness $L_{stl}$ equals 0.3 nm, which is within the range of values for this parameter mentioned in most studies on this subject. Calculated values of the nanobubbles sizes using measured values of zeta potential are shown as a red line on Figure 4. It is seen that theory agrees quite well with experimental values.

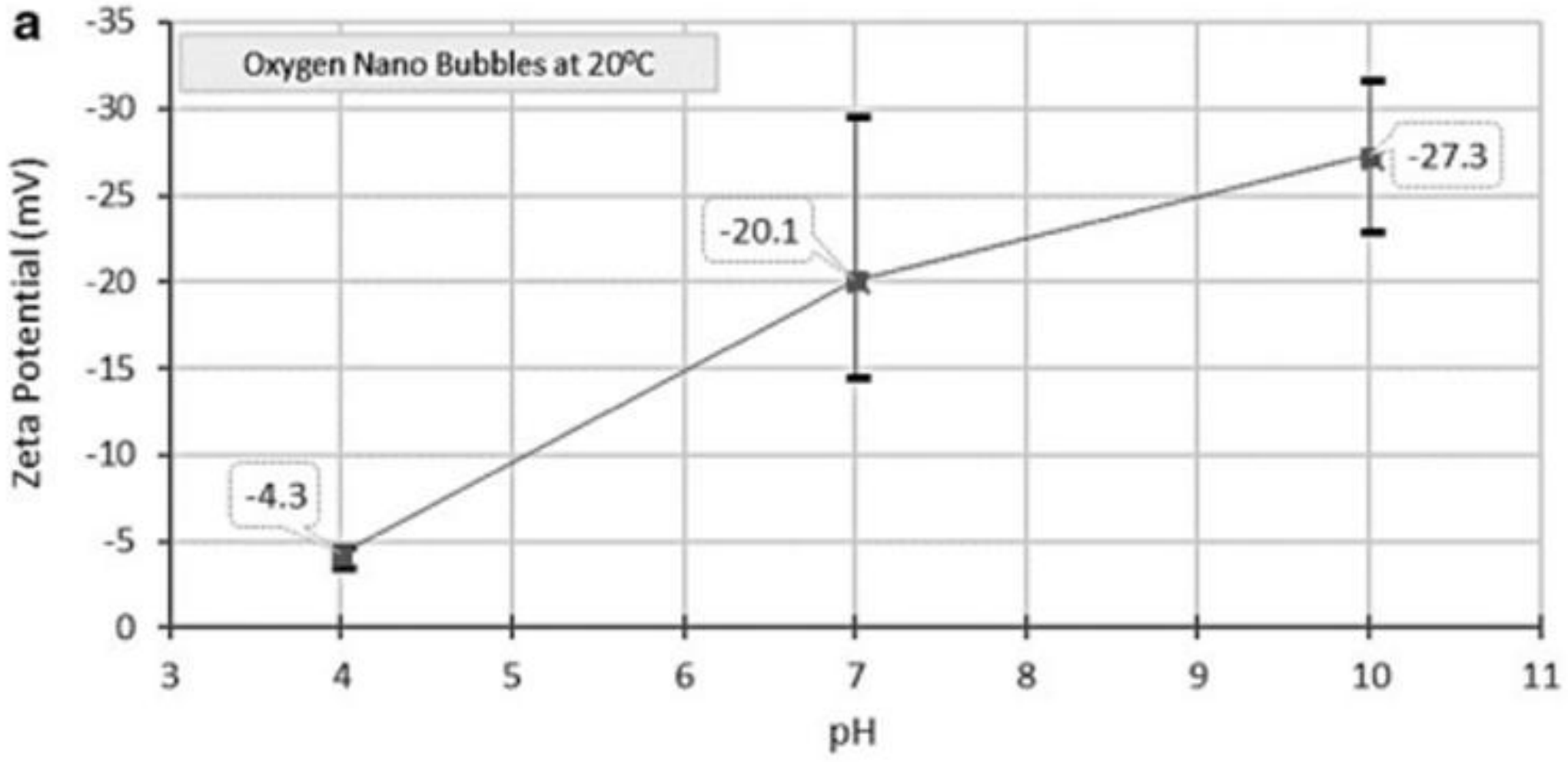

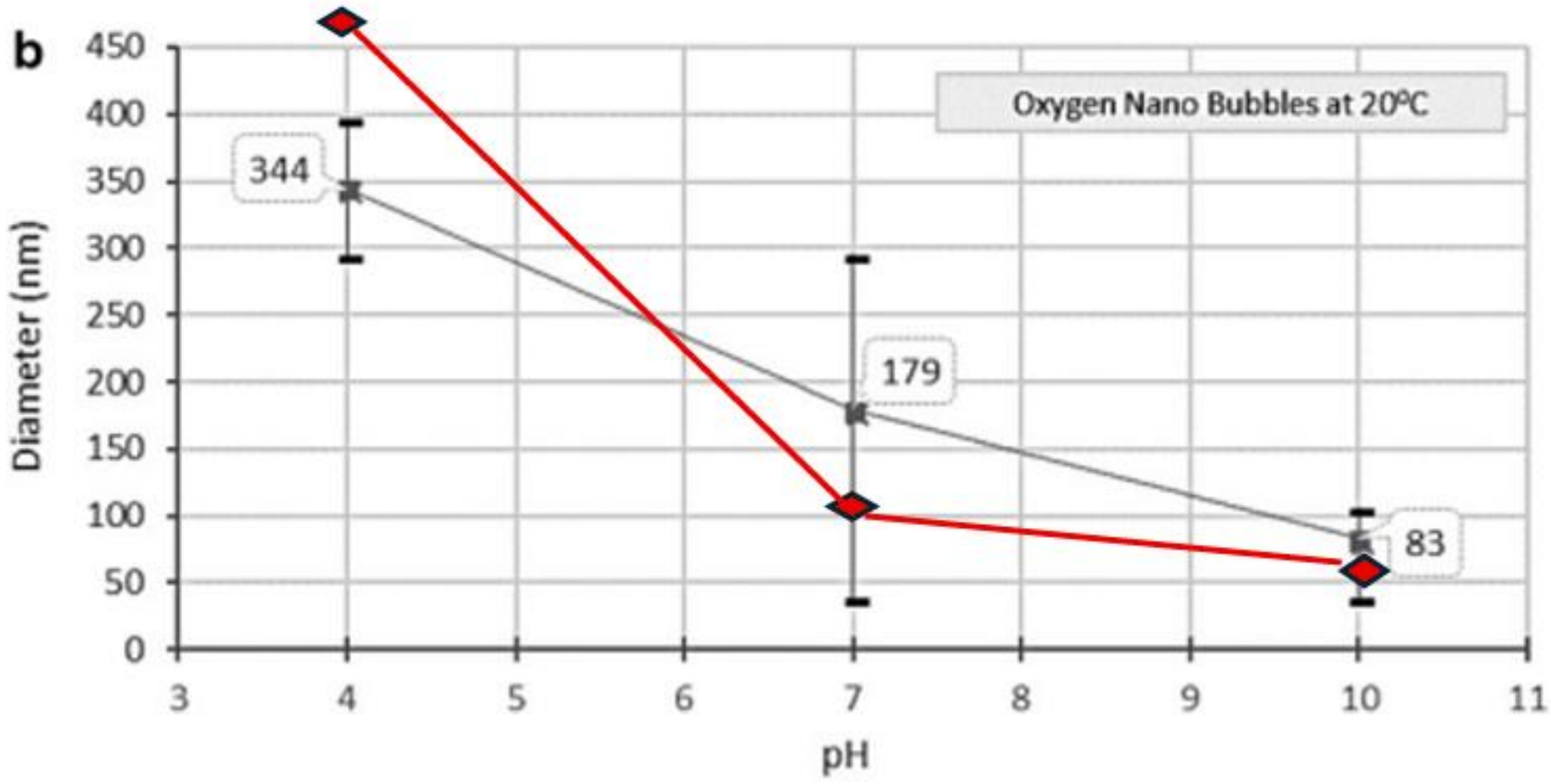

**Figure 4. Relationship between (a) zeta potential and pH and (b) bubble diameter and pH for oxygen nano-bubbles solution at 20$^0$ C. Experimental data is from paper [78]. Red line is theoretical value according to Eq.5, assuming thickness of the structured water layer as 0.3 nm.**

The balance between dielectrostatic force and Young-Laplace pressure is only a necessary condition for the stable nanobubble because it ensures compensation of forces in the normal to the interface direction only. There must be a factor that balances the surface tension in the lateral direction, as was stated in Question 2 above. This factor is "repulsion of oriented water molecules dipole moments", as was suggested in the papers [48,49].

There was equation derived in the paper [48] for the contribution of this repulsion factor to the surface tension:

$$\gamma_{str} = \frac{2 * d_w^2}{\pi \varepsilon \varepsilon_0 L^5} \quad (6)$$

where L is a distance between dipole moments within structured water layer.

The influence of this parameter L is very strong, power 5. Therefore, small variations could lead to large variation in the surface tension effect. There is an estimate of the average distance between water molecules at room temperature – 0.28 nm. However, we should keep in mind that surface tension pushes water molecules closer to each other within the structured layer and distance L is shorter. It turns out that L=0.23 nm leads to complete balance between surface tension and repulsion of the water molecules dipoles in the structured interfacial layer. This value is reasonably close to the value in the water bulk, which also supports this theory.

Figure 5 demonstrates force balance for the stable nanobubble in both directions: normal to interface and the lateral one.

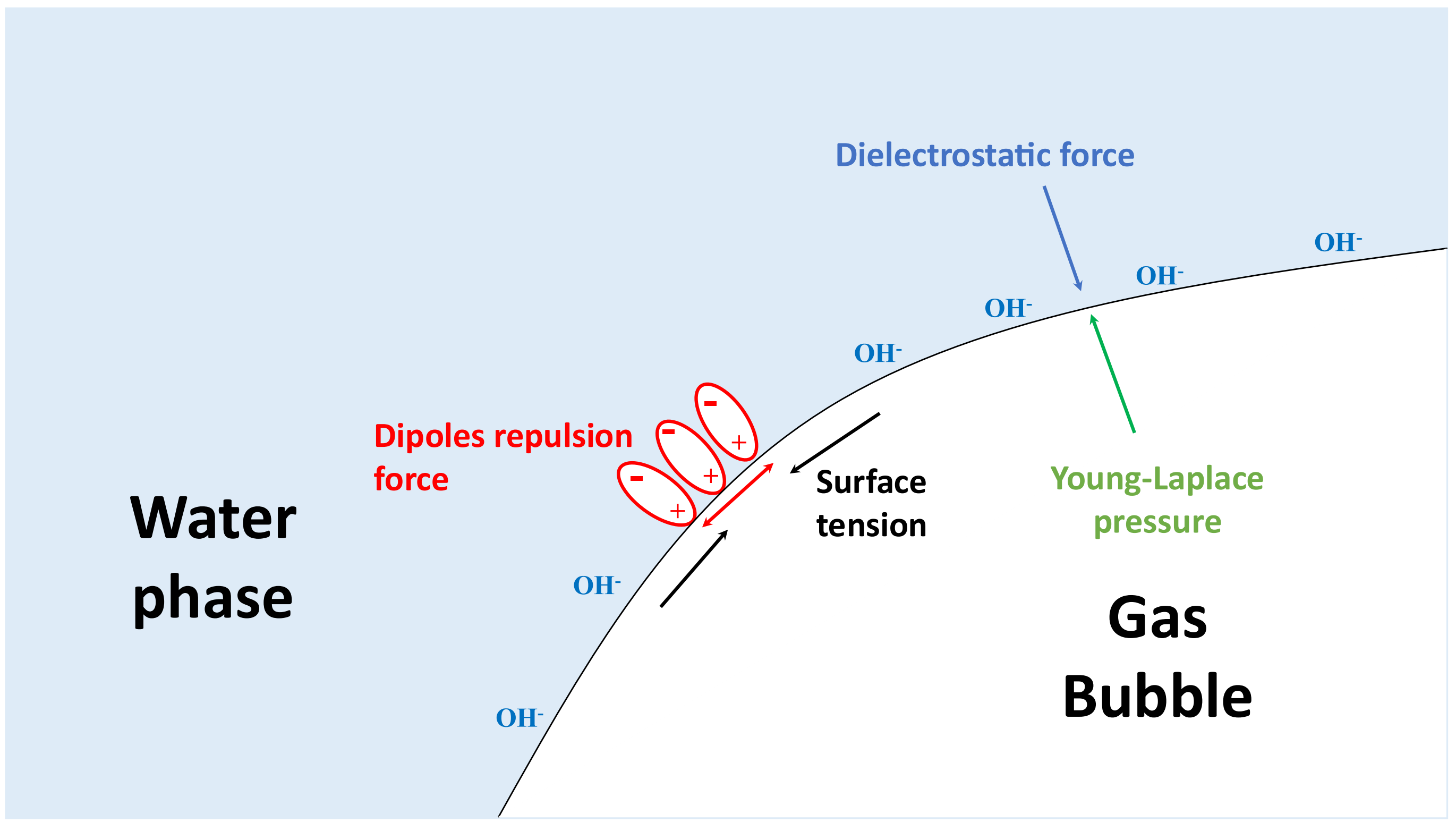

**Figure 5. Illustration of the balance at interface between normal (Young-Laplace and dielectrostatic) forces and lateral (surface tension and repulsion of dipole moments) forces.**

Finally, we can answer fundamental questions formulated at the beginning of this section. Here they are:

1. Interaction of inhomogeneous electric field of EDL with oriented water molecules dipole moments creates a force (dielectrostatic force) that is capable of completely compensating Young-Laplace pressure;
2. Repulsion of the oriented water molecules dipole moments contribution to the surface tension is comparable with classical surface tension and can compensate for its effect on interface;
3. Balance between Young-Laplace pressure and dielectrostatic force determines stable nanobubble size in the range of hundreds of nanometers, according to Eq.5.
4. Expression for the stable nanobubbles size, Eq. 5, predicts reciprocal linear relationship between size and zeta potential, same as in experiment shown in Figure 4.

The same theory could be applied to studying pristine emulsions.

## Pristine emulsions.

The term "pristine emulsions" was introduced in the paper [48] following Beattie and Djerdjev [24] who used term "pristine" for interfaces having no surfactants. These emulsions consist of only two liquids with no added surfactant, and they are different from Pickering emulsions [50], which are stabilized with colloid particles.

Existence of such emulsions contradicts classical physical chemistry model of emulsion that requires presence of adsorption layer for stabilizing liquid-liquid interface according to the paper [51] cited above for nanobubbles. Nevertheless, there is a large bulk of publications dedicated to experimental observations and studies of these emulsions. We have found 22 published papers that are organized below in Table 1.

**Table 1. Summary of publications on the "pristine emulsions". All emulsions consist of water and oil. The oil phase might be different. We list these oils in the third column, which is labeled "oil".**

| Year | Country, affiliation, reference | Oil | Molecular weight |
|---|---|---|---|
| **Oil-in-water emulsions** | | | |
| 1996 | Bulgaria, USA [25] | • Hexadecane<br>• Dodecane<br>• Xylene<br>• Perfluoromethyldecalin | 226.4<br>170.3<br>106.17<br>206.6 |
| 1999-2018 | Japan [26-31] | • Hexadecane<br>• Hexane<br>• Benzene<br>• Oleic acid<br>• Esters of oleic acid | 226.4<br>86.2<br>78.1<br>282.5<br>296-338 |
| 2003-2004 | University of California, USA [23,32,33] | • Dodecane<br>• Hexane<br>• Octane<br>• Decane<br>• Octadecane<br>• Squalane<br>• Hexamethylsqualane<br>• 4-fluorotoluene | 170.3<br>86.2<br>114.2<br>142.3<br>254.5<br>422.8<br>506.9<br>110.1 |
| 2004 | Bristol University, UK [34] | • Dodecane | 170.3 |
| 2004, 2025 | Beattie and Djerdjev from Sydney University, Australia [24,35] | • Hexadecane<br>• Decane<br>• Dodecane<br>• Eicosane<br>• Squalane<br>• Perfluoromethyldecalin | 226.4<br>142.3<br>170.3<br>282.5<br>422.8<br>206.6 |
| 2020 | USA [41] | • Hexadecane | 226.3 |
| 2022 | Korea [36] | • Olive oil (triolein) | 885.5 |
| 2023 | China, Japan, Canada [40] | • Decane | 142.3 |
| 2025 | USA and China [42,43] | • Hexadecane | 226.4 |
| 2018,2025 | France [44-47] | • Hexadecane | 226.4 |
| 2026 | USA [48,49] | • Hexadecane | 226.4 |
| **Water-in-oil emulsion** | | | |
| 2010-2018 | Japan [37,38,39] | • Cyclohexane<br>• Dodecane<br>• Benzene<br>• Octane<br>• Hexane<br>• Oleic acid | 84.2<br>170.3<br>78.1<br>114.2<br>86.2<br>282.5 |

Pristine emulsions are apparently very promising for many applications described in the paper by Maeda et al [33]. In particular, they allow controlling chemical kinetics according to the extensive review by LaCour et al [43]. This review describes examples of many organic and redox reactions that occur with much faster kinetics in water microdroplets and oil–water emulsions than in bulk solution.

There is one serious difference between pristine emulsions and nanobubbles, besides similarity in their pristine interfaces. The outflow diffusional fluxes are not that critical for pristine emulsions compared to nanobubbles. In general, effect on such fluxes on emulsion stability is known as Ostwald ripening [118].

Ostwald ripening is a phenomenon observed in emulsions that involves the change of a droplet size over time due to transfer of emulsion droplet liquid from small droplets into the larger ones. This is a thermodynamically driven process caused by excessive pressure in the emulsion droplets due to curvature of their interfaces. It is usually referred to as Kelvin's pressure [119120,121] and it is similar to Young-Laplace pressure in the case of bubbles [60,61]. This excessive pressure increases chemical potential of the liquid molecules that form emulsion droplets. It becomes higher in the vicinity of smaller droplets compared to the vicinity of the larger droplets. Resulting diffusion flux moves these molecules from small droplets to larger ones. This description stresses the importance of polydispersity for Ostwald ripening [118]. However, it is possible to derive an equation for evolution of an average particle volume assuming some typical droplet size distribution. This was done by Lifshitz, Slyozov and Wagner [122,123] who derived an equation for characterizing kinetics of Ostwald ripening. It yields the following expression for characteristic time of this process $t_{lsw}$ (sec) [48,49]:

$$t_{lsw} = \frac{9\rho^2 RT a_0^3}{8DC_{sol}\gamma M} \qquad (7)$$

where $a_0$ is an average radius of the droplets in [m], D is diffusion coefficient of the emulsion droplet liquid in [$m^2$/sec], $C_{sol}$ is solubility in [$g/m^3$], $\gamma$ is surface tension in [N/m], M is molecular weight in [g/mol], $\rho$ is density in [$g/m^3$], R is gas constant in [$J/K^0$ mol], T is absolute temperature in $K^0$.

This characteristic time of the Ostwald ripening was calculated in the papers [48,49] for 4%vl hexadecane-in-water emulsion, one of the typical examples of such systems. The result is 38 days, which is many orders of magnitude longer than for nanobubbles. Therefore, we can conclude that Ostwald ripening and outflow diffusional fluxes do not affect longevity of pristine emulsion, which distinguishes them from nanobubbles and does not affect observed emulsion evolution.

The other effect that could affect longevity of emulsions is coalescence. In principle, emulsions could be shielded against coalescence by electrostatic stabilization. However, this coalescence is unusual because the emulsion droplet size seems to reach a constant value as was shown for instance in the study [25,48,49]. The observed value of this stable size agrees with the calculated value based on Eq.5. This agreement between experiment and theory confirms role of the structured water layer and two new surface forces (dielectrostic and dipoles moments repulsion) discussed in the previous section.

There is one more aspect that is paradoxical and can be resolved with introduction of these two new surface forces – it is emulsification by itself. Usually, emulsification requires surfactant according to IUPAC report [51]. Surfactants reduce surface tension which promotes creation of the fresh interface and facilitates emulsification. Emulsification without surfactants is unusual and paradoxical. Existence of the structured water later and its interaction with EDL resolves this paradox since water molecules dipole moments repulsion reduces surface tension.

There are still some unanswered questions regarding pristine emulsions. For instance, it is not quite clear yet how droplets reach stable size by redistributing internal liquid phase during collisions, as it was discussed for instance in the paper [124]. Another question is about kinetics of the structured water layer and EDL during emulsification.

One more question is about nature of the oil phase that could create such structured water layer at interface. Not all liquids can form pristine emulsion. For instance, toluene does not create them according to the report [48].

There is a hypothesis that the capability of creating such pristine emulsion is linked simply to the small size of the oil phase molecule. Larger molecules can have a more stable emulsion. The molecular weight of toluene 92 g/mol is much smaller than molecular weight of hexadecane 226 g/mol. This hypothesis is supported by old observation by Frank and Evans [98]. They attributed insolubility of non-polar substances in water more to entropy effect, rather than internal energy. Here what they wrote: …"*Large, non-polar molecules have stronger van der Waals force fields around them than do small ones, and are more strongly held in any condense phase, including aqueous solutions. The larger they are, however, the larger the iceberg which they produce in water, and therefore the greater the loss in entropy involved in dissolving them…".*

In order to verify this hypothesis, we listed molecular weights of oils that presumably form pristine emulsions according to Table 1. Most of the liquids there possess high molecular weights indeed. However, hexane, benzene, and cyclohexene have molecular weight below 100 and less than toluene. This seemingly disapproves of this explanation. It would be prudent in the future to verify existence of pristine emulsions with these oil phases and confirm that impurities do not play any role.

Nevertheless, it seems that introduction of structured water layer interaction with EDL has removed a veal of mystery from these fascinating and potentially very important emulsions.

## Conclusions

The ***structured water layer*** at hydrophobic interfaces was first proposed more than a century ago and has since been supported by many prominent scientists and extensive experimental evidence. It turns out that ***structured water layer*** is essential for resolving three paradoxes in most important fields of Colloid and Interface Science: electrokinetics and stability.

This interfacial layer modifies solubilities of ions at interface, which leads to the charge separation and formation additional component of the electric double layer (EDL). This component of EDL exists at high ionic strength, which explains many observations of the electrokinetic phenomena at the ionic strength range where classical theory denies it.

Interaction of structured water layer with traditional EDL could explain paradoxical longevity of surfactant-free nanobubbles. This interaction creates forces that compensate Young-Laplace pressure in normal to interface direction and surface tension in the lateral direction. Corresponding theoretical model predicts nanobubbles size that agrees with experiment. It also explains experimentally observed reciprocal proportionality between nanobubble size and zeta potential. Basically, this theory explains existence of "stability island" for bubbles in the range of hundreds of nanometers.

The same forces could be invoked to explain existence of pristine emulsions that have no surfactant. It explains why emulsification is possible without surfactant due to their stabilizing action during emulsification. It also explains observed evolution of these emulsions towards stable droplet size. Calculated value of such size agrees with experiment.

## References

1. Gouy M: Sur la constitution de la charge électrique à la surface d'un electrolyte. *J Phys Theor Appl* 1910: 9, 457-468.

2. Chapman DL: LI. A contribution to the theory of electrocapillarity. *Lond Edinb Dublin Philos Mag J Sci* 1913: 25, 475-481.
3. Debye P, Hückel E: Zur Theorie der Elektrolyte. I. Gefrierpunktserniedrigung und verwandte Erscheinungen. *Phys Zeit* 1923:24, 185–206. (Translated by Braus, MJ, The Theory of Electrolytes. I. Freezing Point Depression and Related Phenomenon, 2019).
4. Johnson SB, Scales PJ, Healy TW: The binding of monovalent electrolyte ions on $\alpha$-alumina. 1. Electroacoustic studies at high electrolyte concentrations. *Langmuir* 1999:15, 2836-2843.
5. Kosmulski M, Rosenholm JB: High ionic strength electrokinetics. *Adv Colloid Interfac* 2004:112, 93-107.
6. Kosmulski M, Dukhin AS, Priester T, Rosenholm JB: Multi laboratory study of the shifts in the IEP of anatase at high ionic strengths. *J Colloid Interf Sci* 2003:263, 152–155.
7. Kosmulski M, Dahlsten P: High ionic strength electrokinetics of clay minerals. *Colloid Surface A* 2006:291, 212-218.
8. Kosmulski M, Dahlsten P, Próchniak P, Rosenholm, JB: High ionic strength electrokinetics. *Colloid Surface A* 2007:301, 425-431.
9. Dukhin AS, Dukhin SS, Goetz PJ: Electrokinetics at high ionic strength and hypothesis of the double layer with zero surface charge. *Langmuir* 2005:21, 9990-9997.
10. Garg A, Cartier CA, Bishop KJM, Velegol D: Particle zeta potentials remain finite in saturated salt solutions. *Langmuir* 2016:32, 11837−11844.
11. Coday BD, Luxbacher T, Childress AE, Almaraz N, Xu P, Cath TY: Indirect determination of zeta potential at high ionic strength. Specific application to semi-permeable polymeric membranes. *J Membrane Sci* 2015:478, 58-64.
12. Bikerman JJ: XXXIX. Structure and capacity of electrical double layer. *Lond Edinb Dublin Philos Mag J Sci* 1942: 33, 384-397.
13. Freise V: Zur theorie der diffusen Doppelschicht, *Zeitschrift für Elektrochemie, Berichte der Bunsengesellschaft für physikalische Chemie* 1952: 56, 822-827.
14. Kornyshev AA: Double-layer in ionic liquids: Paradigm change? *J Phys Chem B* 2007: 111, 5545-5557.
15. Bazant MZ, Storey BD, Kornyshev AA: Double layer in ionic liquids: Overscreening versus crowding. *Phys Rev Lett* 2011:106, 046102.

16. Goodwin ZA, Kornyshev AA.: Underscreening, Over screening and Double-Layer Capacitance. *Electrochem. Commun*. 2017, 82, 129−133
17. Lyklema, J, Fundamentals of Interface and Colloid Science, Volumes 1, *Academic Press*, 1993.
18. Carnie SL, Torrie GM: The statistical mechanics of the electrical double layer. In Prigogine I, Rice SA, editors: *Advances in Chemical Physics,* New York, 1984, John Wiley & Sons pp142-253.
19. Mancui M, Ruckenstein E: The polarization model for hydration/double layer interactions: the role of the electrolyte ions. *Adv Colloid Interfac* 2004:112, 109-128.
20. Arias FJ, "Comment on the paper 'A new approach to explaining nano-bubbles paradoxical longevity' Dukhin A and Xu R [Colloids Surf A 700 (2024) 34805", Colloids and Surfaces A, 727, 2, 137962, 2025
21. Yaroshchuk AE., A model for shift of isoelectric point of oxides in concenetred and mixed-solven electrolyte solution, JCIS, 238, 381-384, 2001
22. Eastoe J, Ellis C. De-gassed water and surfactant-free emulsion. History, controversy, and possible application. Adv in Colloid and Interface Sci., 134-135, 89-95, 2007.
23. Pashley RM. Process for the production of emulsions and dispersions., US Patent 7,696,252 B2, 2010
24. Beattie JK and Djerdjev AM, "The Pristine Oil/Water Interface: Surfactant-Free Hydroxide-Charged Emulsions", Angew. Chem. Int. Ed. 43,3568–3571, 2004.
25. Marinova KG, Alargova RG, Denkov ND, Velev OD, Petsev DN, Ivanov IB, Borwankar RP. Charging of oil-water interfaces due to spontaneous adsorption of hydroxyl ions. Langmuir, 12, 2045-2051, 1996.
26. Kamogawa K, Matsumoto M, Kobayashi T, Sakai T, Sakai H, Abe M, Langmuir, 15, 1913-1919, 1999.
27. Sakai T, Kamogawa K, Harusawas F, Momozawa N, Sakai H, Abe M, "Direct observation of flocculation/coalescence of metastable oil droplets in surfactant-free oil/water emulsion by freeze-fracture electron microscopy", Langmuir, 17 (2), 255-259, 2001.
28. Kamogawa K, Akatsuka H, Matsumoto M, Yokoyama S, Sakai T, Sakai H, Abe M, "Surfactant-free o/w emulsion formation of oleic acid and its esters with ultrasonic dispersion", Colloids and Surfaces, A, 180, 41-53, 2001.

29. Sakai T, Kamogawa K, Nishiama K, Sakai H, Abe M, "Molecular diffusion of oil/water emulsions in surfactant-free conditions", Langmuir, 18 (6), 1985-1990, 2002.
30. Sakai T, Takeda Y, Mafune F, Abe M, Kondow T. "Monitoring growth of surfactant-free nanodroplets dispersed in water by single-droplet detection", J. Phys. Chem., 107 (13), 2921-2926, 2003.
31. Sakai T., "Surfactant-free emulsions", Current Opinion Colloid Interface Science, 13(4), 228-235, 2008.
32. Pashley RM. "Effect of degassing on the formation and stability of surfactant-free emulsions and fine Teflon dispersions", J. Phys. Chem., B 107, 1714-1720, 2003.
33. Maeda N, Rosenberg KJ, Israelachvili JN, Pashley RM. "Further studies on the effect of degassing on the dispersion and stability of surfactant-free emulsions", Langmuir, 20, 3129-3137, 2004.
34. Burnett GR, Atkin R, Hicks S, Eastoe J. "Surfactant-free "emulsions" generated by freeze-thaw", Langmuir, 20, 5673-5678, 2004.
35. Djerdjev AM, and Beattie JK, "Hydroxide and hydrophobic terabutylammonium ions at the hydrophobe-water interface", Molecules, 30, 875, 2025
36. Hwanbo SAH, Lee SY, Kim BA, Moon CK, "Preparation of surfactant-free nano oil particles in water using ultrasonic system and mechanism of emulsion stability", Nanomaterials, 12, 1547, 2022.
37. Horikoshi S., Akao Yu, Ogura T, Sakai H, Abe M, Serpone N. "On the stability of surfactant-free water-in-oil emulsions and synthesis of hollow $SiO_2$ nanospheres", Coll. Surf. A, 372, 55-60, 2010.
38. Sakai T, Seo K. "Colloidal stability of emulsifier-free water-in-oil emulsions: effect of oil property", J. Japan Soc Colloid Mater, 87(11), 1-6, 2014.
39. Sakai T, Oishi T. "Colloidal stabilization of surfactant-free emulsion by control of molecular diffusion among droplets", J. Taiwan Inst. Chem. Engin., 92, 123-128, 2018.
40. Alizadeh A, Huang Y, Liu F, Daiguji H, Wang M. "A streaming-potential-based microfluidic measurement of surface charge at immiscible liquid-liquid interface", Int. J. Mech. Sci., 247, 108200, 2023
41. Xiong H, Lee JK, Zare RN, Min W. "Strong electric field observed at the interface of aqueous microdroplets", The Journal of Physical Chemistry Letters, 11, 7423-7428, 2020.

42. Shi L, LaCour RA, Oian N, Heindel JP, Lang X, Zhao R, Head-Gordon T, Min W. "Water structure and electric fields at the interface of oil droplets", Nature, 640, 87-93, 2025.
43. LaCour RA, Heindel JP, Zhao R, G, Head-Gordon T. "The role of interfaces and charge for chemical reactivity in microdroplets", Journal of American Chemical Society, 147, 6299-6317, 2025.
44. Ganachaud F., Sauzedde F, Elaissari A, Pichot C. "Emulsifier-free emulsion copolymerization of styrene with two different amino-containing cationic monomers. 1. Kinetic studies." The Journal of Applied Polymer Science, 65, 12, 2315-2330, 1997.
45. Ganachaud F. "An alternative hypothesis of enhanced deep supercooling of water: nucleator inhibition via bicarbonate adsorption", The Journal of Physical Chemistry Letters, 16, 261-264, 2025.
46. Yan X, Delgado M, Aubry J, Gribelin O, Stocco A, Cruz FBD, Bernard J, Ganachaud F. "Central role of bicarbonate anions in charging water/hydrophobic interfaces", The Journal of Physical Chemistry Letters, 9, 96-103, 2018.
47. Yan X, Stocco A, Bernard J, Ganachaud F. "Freeze/thaw-induced carbon dioxide trapping promotes emulsification of oil in water", The Journal of Physical Chemistry Letters, 9, 5998-6002, 2018.
48. Dukhin AS, Xu R, Velegol D, "Evolution of pristine emulsions", https://arxiv.org/pdf/2509.13529
49. Dukhin A.S., Xu Renliang, Velegol D., "Evolution of pristine emulsions and hypothesis explaining their existence", Int.J. Molecular Science, 27, 1837, 2026.
50. Pickering, Spencer Umfreville "Emulsions". Journal of the Chemical Society, Transactions,91, 2001–2021, 1907
51. "Measurement and Interpretation of Electrokinetic Phenomena, International Union of Pure and Applied Chemistry, Technical Report", *Pure Appl Chem*, 77(10), 1753-1805, 2005
52. Snoeyink VL, Jenkins D, *Water Chemistry*, John Wiley & Sons, NY, 1980.
53. Derjaguin BV, Landau L: Theory of the stability of strongly charged lyophobic sols and the adhesion of strongly charged particles in solution of electrolytes. *Acta Phys Chim USSR* 14: 733, 1941.
54. Verwey EJW, Overbeek JThG: *Theory of the Stability of Lyophobic Colloids.* Amsterdam, 1948, Elsevier.

55. Liftshitz EM: Theory of molecular attractive forces. *Soviet Phys JETP* 2: 73, 1956.
56. Israelachvili JN, Pashley RM, The hydrophobic interaction is long range, decaying exponentially with distance. Nature, 300, 341-342, 1982
57. Dukhin AS, Xu R., A new approach to explain nano-bubbles paradoxical longevity, Colloids and Surfaces, A, 700, 134805, 2024.
58. Epstein PS, Plesset MS. On the stability of gas bubbles in liquid-gas solutions, J. Chem. Phys., 18, 1505-1509, 1950
59. Ljunggren S, Eriksoon JC: The lifetime of a colloid-sized gas bubble in water and cause of hydrophobic attraction. *Colloid Surface A* 1997:129, 151-155.
60. Young, T., "An Essay on the Cohesion of Fluids", *Philos T R Soc Lond*, 95, 65–87, (1805)
61. marquis de Laplace, P.S., Traité de Mécanique Céleste, (Paris, France: Courcier), *Supplément au dixième livre du Traité de Mécanique Céleste*, 4, 1–79, (1805)
62. Foudas AW, Kosheleva RI, Favvas EP, Kostoglou M, Mitropoulos AC, Kyzas GZ, Fundamentals and Applications of Nanobubbles: A Review, *Chem Eng Res Des*, 189, 64-86, 2023
63. Michailidi ED, Bomis G, Varoutoglou A, Efthimiadou EK, Mitropoulos AC, Favvas, EP. Fundamentals and applications of nanobubbles. *Advanced Low-Cost Separation Techniques in Interface Science*. Interface Science and Technology. 30, 69–99, 2019
64. Azevedo A, Etchepare R, Calgaroto S, Rubio J. Aqueous dispersions of nanobubbles: Generation, properties and features. Minerals Engineering. **94**: 29–37, 2016
65. Nirmalkar N, Pacek AW, Barigou M. On the Existence and Stability of Bulk Nanobubbles. *Langmuir*. **34** (37): 10964–10973, 2018
66. Alheshibri M, Qian J, Jehannin M, Craig VS, A History of Nanobubbles, *Langmuir*, 32(43), 11086-11100, 2016
67. Babu KS, Amamcharla JK, Generation Methods, Stability, Detection Techniques, and Applications of Bulk Nanobubbles in Agro-Food Industries: A Review and Future Perspective, *Critical Reviews in Food Science and Nutrition*, pp1-20, Taylor & Francis, 2002
68. Wang Q, Zhao H, Qi N, Qin Y, Zhang X, Li Y, Generation and stability of size-adjustable bulk nanobubbles based on periodic pressure change, *Nature, Scientific reports*, 9, 1118, 2019

69. Atkinson AJ, Apul OG, Schneider O, Garcia-Segura S, Westerhoff P. Nanobubble Technologies Offer Opportunities to Improve Water Treatment. *Accounts of Chemical Research*. **52** (5): 1196–1205. 2019
70. Temesgen T, Bui TT, Han M, Kim T, Park H. Micro and nanobubble technologies as a new horizon for water-treatment techniques: A review. *Advances in Colloid and Interface Science*. **246**: 40–51, 2017
71. Wu J, Zhang K, Cen C, Wu X, Mao R, Zheng Y. Role of bulk nanobubbles in removing organic pollutants in wastewater treatment. *AMB Express*. **11** (1): 96., 2021
72. Nazari S, Hassanzadeh A, He Y, Khoshdast H, Kowalczuk PB. Recent Developments in Generation, Detection and Application of Nanobubbles in Flotation. Minerals. **12** (4): 462, 2022
73. Lyu T, Wu S, Mortimer RJG, Pan G. Nanobubble Technology in Environmental Engineering: Revolutionization Potential and Challenges. E*nvironmental Science & Technology*. **53** (13): 7175–7176, 2019
74. Zhou Y, Han Z, He C, Feng Q, Wang K, Wang Y, Luo N, Dodbiba G, Wei Y, Otsuki A, Fujita T, Micro-Nanobubble Technology in Surface Cleaning and Defouling. Applied Sciences. **15** (3): 1197., 2025
75. Liu S, Enari Y, Kawagoe Y, S, Oshita M, Properties of the water containing nanobubbles as a new technology of the acceleration of physiological activity, Elsevier, Chem. Eng. Sci, 93, 250-256, 2012
76. Oshita S., Liu S, Nanobubble characteristics and its application to agriculture and food, Int. Symp. Agri-Foods Healing Wealth, 2, 23-32, 2013
77. Lee JI, Kim JM, Influence of temperature on bulk nanobubbles generation by ultrasound, Colloid and Interface Science Communications, 49, 100639, 2022
78. Meegoda JN, Hewage SA, Batagoda JH, Stability of Nanobubbles, *Environ Eng Sci*, 35(11), 1216-1227, 2018
79. Hewage SA, Kewalramani J, Meegoda JN, Stability of Nanobubbles in Different Salts Solutions, *Colloid Surface A*, 609, 125669, 2021
80. Hewage SA, Meegoda JN. (2022). Molecular dynamics simulation of bulk nanobubbles. *Colloids and Surfaces A: Physicochemical and Engineering Aspects*. **650** 129565, 2022

81. Karimi, Mahammadjavad; Parsafar, Gholamabbas; Samouei, Hamidreaza, "Polarizing Perspective: Ion- and Dipole-Dipole dipole Interactions Dicate Bulk Nanobubble Stability", *J. Phys.Chem.*, **128**: 7263–7270, 2024
82. Dockar D, Sullivan P, Mifsud J, Gibelli L, Borg MK. Ionic adsorption on bulk nanobubbles interfaces and its uncertain role in diffusive stability. JCIS, 695, 137747, 2025
83. Ma X, Li M, Xu X, Sun C, On the role of surface charge and surface tension tuned by surfactant in stabilizing bulk nanobubbles, Appl. Surf. Sci., 608, 155232, 2023
84. Ma X, Li M, Pfeiffer P, Eisener J, Ohl CD, Sun C, Ion adsorption stabilizes bulk nanobubbles, JCIS, 606, 1380-1394, 2022
85. Tan BH, An H, Ohl CD. How bulk nanobubbles might survive, Phys. Rev. Lett, 124, 134503, 2020
86. Gracia A, Creux P, Lachaise J, Electrokinetics of Gas Bubbles, in Interfacial Electrokinetics and Electrophoresis, ed. Delgado, Á.V., Chapter 29, 825-836, CRC Press, 2002
87. Li, C., Somasundaran, P., “Reversal of Bubble Charge in Multivalent Inorganic Salt Solutions--Effect of Magnesium”, *J Colloid Interface Science*, 146, 215-218, (1991)
88. Ninham, B.W., Yaminsky,V., “Ion Binding and Ion Specificity: The Hofmeister Effect and Onsager and Lifshitz Theories”, *Langmuir*, 13, 2097-2108, (1997)
89. Craig VSJ, Ninham BW, Pashley RM. The effect of electrolytes on bubble coalescence in water. J Phys Chem 1993;97:10192–7. https://doi.org/10.1021/j100141a047.
90. Craig VSJ, Ninham BW, and Pashley RM, “Direct measurement of hydrophobic forces: a study of dissolved gas, approach rate, and neutron irradiation”, Langmuir, 15, 1913-1919, 1999.
91. Ninham BW, Pashley RM, Nostro PL. Surface forces: Changing concepts and complexity with dissolved gas, bubbles, salt and heat. Curr Opinion Colloid In 2017;27:25–32.
92. Ninham BW, Nostro PL. Unexpected Properties of Degassed Solutions. J Phys Chem B 2020;124:7872–8.
93. Onsager L., Samaras, N.N.T., “The Surface Tension of Debye-Hückel Electrolytes”, *J Chem Phys*, 2, 628, 1934
94. Jadhav AJ, Barigou M, On the clustering of nanobubbles and their colloidal stability, JCIS, 601, 816-824, 2021
95. Levin Yu.K., Effect of hydrogen bonds on the stability of nanobubbles in water, Nanostructures and nanotubes, 19, 568-571, 2024

96. Walker-Gibbons R, Kubincova A, Hunenberger PH, Krishan M. The role of surface chemistry in the orientational behavior of water at interfaces, J. Physical Chemistry, B, 126, 4697-4710, 2022
97. Hardy WB: An application of the principle of dynamical similitude to molecular physics. *P R Soc Lond A-Cont* 92, 82-100, 1915.
98. Frank HS and Evans MW, Free volume and entropy in condensed systems. III. Entropy in binary liquid mixtures; Partial molal entropy in dilute solutions; structure and thermodynamics in aqueous electrolytes. Journal of Chemical Physics, 13, 11, 507-532, 1945.
99. Bockris JO'M, Devanathan MAV, Müller K, On the Structure of Charged Interfaces, *Proc R Soc Lon Ser-A*, 274 (1356), 55–79, 1963.
100. Hunter RJ, Zeta Potential in Colloid Science, *Academic Press*, 1986.
101. Pashley RM, Israelachvili JN, Molecular layering of water in thin films between mica surfaces and its relation to hydration forces, *J Colloid Interf Sci*, 101(2), 500-514, 1984.
102. Xiong H, Lee JK, Zare RN, Min W. "Strong electric field observed at the interface of aqueous microdroplets", The Journal of Physical Chemistry Letters, 11, 7423-7428, 2020.
103. Zobel M, Neder R, Kimber SA, Universal solvent restructuring induced by colloidal particles, Science, 347, 6219, 292-294, 2015
104. Tyrode, E.; Liljeblad, J. F. Water structure next to ordered and disordered hydrophobic silane monolayers: a vibrational sum frequency spectroscopy study. J. Phys. Chem. C , 117, 2013
105. Derjaguin BV, Karasev V, and Khromova E., Thermal expansion of water in fine pores, J. Colloid Interface Sci. 109, 586-587, 1986.
106. Ogbebor J, Valenza JJ, Ravikovitch PI, Karunarathe A, Muraro G, Lebedev M, Gurevich B, Khalizov AF, G GY. Ultrasonic study of water adsorbed in nanoporous glasses, Physical Review, E108, 024802, 2023
107. Wang Z., Confined water in low-dimensional materials: structural, spectroscopic, and transport phenomena. Academia NANO: Science, Materials, Technology., 3, 1-23, 2026
108. O'Brien RW, Midmore BR, Lamb A, Hunter RJ: Electroacoustic studies of moderately concentrated colloidal suspensions. *Faraday Discuss Chem Soc* 1990:90, 1-11.
109. O'Brien RW, Cannon DW Rowlands WN: Electroacoustic determination of particle size and zeta potential. *J Colloid Interf Sci* 1995:173, 406-418.

110. Dukhin AS, Goetz JP: *Characterization of Liquids, Nano- and Microparticulates, and Porous Bodies using Ultrasound Ed 3*, London, Elsevier, 2017.

111. Dukhin AS and Xu Renliang, Zeta Potential. Fundamentals, Methods, Applications, Elsevier, 2025

112. Liu Y, Zhou L, Zhang L, Hu J,Ozone Micro–Nanobubbles: Properties, Effects, and Applications, Water, 18(10):1189, 2026

113. Arias FJ, A first theoretical assessment of hydrogen nanobubble generation for ocean wave energy storage, Journal of Energy Storage, 170, 1227592026, 2026

114. Arias FJ, On the maximum electrical power efficiency of a hydroelectric nanobubble-driven cell, Journal of Colloid and Interface Science, 723, 140954, 2026

115. Xin Y, Yang H, Wang H, Efficient and tunable fabrication of hollow vaterite microspheres by micro-nano bubbles and mechanism study, Journal of Environmental Chemical Engineering, 14, 3, 1224402026, 2026

116. Quierre RF, Sacdalan JPC, Postharvest Pathogens in Strawberry (Fragaria x ananassa): Potential of Plasma-Activated Water and Micro-Nano Bubbles for Control–A Review, Applied Science and Engineering, 19, 2, 2026

117. Pohl HA: *Dielectrophoresis: The Behavior of Neutral Matter in Nonuniform Electric Fields.* Cambridge, Cambridge University Press, 1978

118. Ostwald W. "Studien über die Bildung und Umwandlung fester Körper" [Studies on the formation and transformation of solid bodies] (PDF). Zeitschrift für Physikalische Chemise. **22**: 289–33 1897

119. Sir William Thomson, "On the equilibrium of vapor at a curved surface of liquid", *Philosophical Magazine*, series 4, **42** (282), 448-452, 1871.

120. Thomson JJ, Applications of Dynamics to Physics and Chemistry, 1st ed., Cambridge University Press, London, 1888.

121. Thomson JJ, Conduction of Electricity through Gases, Cambridge University Press, London, 1906.

122. Lifshitz IM, Slyozov VV. The Kinetics of Precipitation from Supersaturated Solid Solutions. Journal of Physics and Chemistry of Solids. **19** (1–2): 35–50, 1961

123. Wagner C. Theory of the aging of precipitates by dissolution-reprecipitation (Ostwald ripening). Zeitschrift für Elektrochemie. **65** (7): 581–591. 1961.

124. Taisne L, Walstra P, Cabane B. Transfer of oil between emulsion droplets, JCIS, 184, 378-390, 1996